# Computational Method for a Fractional Model of the Helium Burning Network

Mohamed I. Nouh[1,2]

[1]Department of Physics, College of Science, Northern Border University, Arar, Saudi Arabia.

Email: nouh@nbu.edu.sa, abdo_nouh@hotmail.com

[2]Department of Astronomy, National research Institute of Astronomy and Geophysics, 11421 Helwan, Cairo, Egypt

**Abstract**: Stellar cores may be considered as a nuclear reactor that play important role in injecting new synthesized elements in the interstellar medium Helium burning is an important stage that contribute to the synthesis of key elements such as carbon, through the triple- α process, and oxygen. In the present paper, we introduce a computational method for the fractional model of the nuclear helium burning in stellar cores. The system of fractional differential equations is solved simultaneously using series expansion method. The calculations are performed in the sense of modified Riemann-Liouville fractional derivative. Analytic expressions are obtained for the abundance of each element as a function of time. Comparing the abundances calculated at the fractional parameter $\alpha = 1$, which represents the integer solution, with the numerical solution revealed a good agreement with maximum error $\varepsilon = 0.003$. The product abundances are calculated at $\alpha = 0.25, 0.5, 0.75$ to declare the effects of changing the fractional parameters on the calculations.

1. **Introduction**

One of the most important problems in astrophysics is the nucleosynthesis in the stellar cores. Hydrogen burning, helium burning, CNO cycle are the main stages that the stars produce in their nuclear reactors, Clayton (1983). The basic equations of the burning networks are differential equations that may be solved simultaneously by numerical or analytical methods, Duroah and Kushwaha (1963), Clayton (1983), Hix and Thielemann (1999) and Nouh et al.

(2003). This system of equations may be called the integer version of the burning models, and the general form could be called the fractional models.

In most stars, energies are derived from the conversion of hydrogen into helium. As a result, the temperature goes high (about 100 million degrees) and the hydrogen fusion starts, the stellar core contracts gravitationally until the temperature is high enough to start helium burning. This set of reactions is called the triple alpha process. Helium brining occurs come into two categories, the first is the hydrostatic burning and the second is the explosive burning.

Nowadays, generalizations of some of the basic differential equations in physics led to new and more deep insight to the macroscopic phenomena in a wide range of areas like anomalous diffusion, signal processing and quantum mechanics (Podlubny 1999; Sokolov et al. 2002; Kilbas et al. 2006; Laskin 2000). In astrophysics, El-Nabulsi (2011) considered a fractional equation of state for white dwarf stars; Bayian and Krisch (2015) have investigated the fractional versions of the stellar structure equations for non-radiating spherical objects, Abdel-Salam and Nouh (2016) and Nouh and Abdel-Salam (2017) introduced a series solution to the fractional isothermal and polytropic gas spheres and obtained a solution converges to the surface of the sphere with a few series terms when compared with the solution of the integer one.

In the present paper we are going to introduce a computational method to model the helium burning network in stellar core. The system of the fractional differential equations will be solved by power series expansion. We derive general recurrence relation for the series coefficients in terms of the fractional parameter and what is called the fractal index. The calculations will be performed in the sense of modified Riemann-Liouville fractional Derivative. Comparison with the solution of the integer version is presented to declare the effects of the fractional parameter on the calculations as well as the synthesis process. The paper is organized as follows. In section 2 we introduce the helium burning model. Section 3 is devoted to the series solution of the system of the differential equations. In section 4 we

2. **Helium Burning Network**

In stellar helium core, a triple alpha collision occurs at temperature of the order of 100 million degrees and the product is $C^{12}$. Then $C^{12}$ captures $He^4$ to form $O^{16}$ which in turn captures $He^4$ to form $Ne^{20}$ and so on, Nouh et al. (2003). These reactions could be listed as

$$3He^4 \rightarrow C^{12} + \gamma + 7.281 Mev$$
$$C^{12} + He^4 \rightarrow O^{16} + \gamma + 7.150 Mev$$
$$O^{16} + He^4 \rightarrow Ne^{20} + \gamma + 4.750 Mev$$

By considering the above reactions, Clayton (1983) set up the following model for the helium burning phase. If x, y, z and r are the helium, carbon, oxygen and neon number of atoms per unit mass of stellar material. The next four equations govern the time dependent change of the abundance as

$$dx/dt = -3ax^3 - bxy - cxz,$$
$$dy/dt = ax^3 - bxy,$$
$$dz/dt = bxy - cxz, \qquad (1)$$
$$dr/dt = cxz.$$

where a, b, and c are the reaction rates in unites of $cm^3\ s^{-1}$.
The factor 3 appears in the equation because three helium nuclei join to form one carbon nucleus.

The system of differential equations (Equation (1)) could be solved by many analytical or numerical methods. The accuracy of the methods of numerical integrations depend mainly on the interval step used through the solution (time step in our case). Bad choice of this time step may lead to bad accuracy as well as numerical instability. This difficulty could be dimensioned if we use analytical methods. In the present article we will implement power series expansion to solve the fractional helium burning network.

### 3. Series Solution of the System of the Fractional Differential Equations
#### 4.1. Series Expressions for the Unknowns

The system of differential equations (Equation (1)) could be written in the fractional form as

$$D_t^\alpha x = -3a\,x^3 - b\,x\,y - c\,x\,z,$$
$$D_t^\alpha y = a\,x^3 - b\,x\,y,$$
$$D_t^\alpha z = b\,x\,y - c\,x\,z, \qquad (11)$$
$$D_t^\alpha r = c\,x\,z,$$

Assuming the transform $T = t^\alpha$, the solution can be expressed in a series form as

$$x = \sum_{m=0}^{\infty} X_m T^m = X_0 + X_1 T + X_2 T^2 + X_3 T^3 + ...$$
$$= X_0 + X_1 t^\alpha + X_2 t^{2\alpha} + X_3 t^{3\alpha} + ...,$$

$$y = \sum_{m=0}^{\infty} Y_m T^m = Y_0 + Y_1 T + Y_2 T^2 + Y_3 T^3 + ...$$
$$= Y_0 + Y_1 t^\alpha + Y_2 t^{2\alpha} + Y_3 t^{3\alpha} + ...,$$

$$z = \sum_{m=0}^{\infty} Z_m T^m = Z_0 + Z_1 T + Z_2 T^2 + Z_3 T^3 + ...$$
$$= Z_0 + Z_1 t^\alpha + Z_2 t^{2\alpha} + Z_3 t^{3\alpha} + ...,$$

$$r = \sum_{m=0}^{\infty} R_m T^m = R_0 + R_1 T + R_2 T^2 + R_3 T^3 + ...$$
$$= R_0 + R_1 t^\alpha + R_2 t^{2\alpha} + R_3 t^{3\alpha} + ....$$

(12)

where $X_m, Y_m, Z_m, R_m$ are constants to be determined.

### 4.2. Fractional derivative of the function raised to powers

The left hand side of the system in Equation (1) represents the abundances of the elements where the helium abundance (x) is raised to power 3. To obtain the fractional derivative of $u^n$ we apply the fractional derivative of the product two functions. Taking the fractional derivative of both sides of Equation (11), we have

$$D_x^\alpha u^n = D_x^\alpha G \Rightarrow n u^{n-1} D_x^\alpha u = D_x^\alpha G$$

or

$$n\, u^n D_x^\alpha u = u\, D_x^\alpha G \qquad (13)$$

Differentiating both sides of Equation (13) $k$ times $\alpha$-derivatives we have

$$\underbrace{D_x^\alpha ... D_x^\alpha}_{k\, times} [nG D_x^\alpha u] = \underbrace{D_x^\alpha ... D_x^\alpha}_{k\, times} (u D_x^\alpha G)$$

or

$$n \sum_{j=0}^{k} \binom{k}{j} \underbrace{D_x^\alpha ... D_x^\alpha}_{j+1 times} u \underbrace{D_x^\alpha ... D_x^\alpha}_{k-j times} G = \sum_{j=0}^{k} \binom{k}{j} \underbrace{D_x^\alpha ... D_x^\alpha}_{j+1 times} G \underbrace{D_x^\alpha ... D_x^\alpha}_{k-j times} u.$$

At $x = 0$, we have

$$n \sum_{j=0}^{k} \binom{k}{j} \underbrace{D_x^\alpha ... D_x^\alpha}_{j+1 times} u(0) \underbrace{D_x^\alpha ... D_x^\alpha}_{k-j times} G(0) = \sum_{j=0}^{k} \binom{k}{j} \underbrace{D_x^\alpha ... D_x^\alpha}_{j+1 times} G(0) \underbrace{D_x^\alpha ... D_x^\alpha}_{k-j times} u(0) \qquad (14)$$

since

$$\underbrace{D_x^\alpha ... D_x^\alpha}_{j+1\ times} u(0) = X_{j+1}\Gamma((j+1)\alpha+1),$$

$$\underbrace{D_x^\alpha ... D_x^\alpha}_{k-j\ times} G(0) = Q_{k-j}\Gamma(\alpha(k-j)+1),$$

$$\underbrace{D_x^\alpha ... D_x^\alpha}_{j+1\ times} G(0) = Q_{j+1}\Gamma((j+1)\alpha+1),$$  (15)

$$\underbrace{D_x^\alpha ... D_x^\alpha}_{k-j\ times} u(0) = X_{k-j}\Gamma(\alpha(k-j)).$$

## 4.2. Recurrence Relations

substituting Equation (14) in Equation (15) we have

$$n\sum_{j=0}^{k}\binom{k}{j}X_{j+1}\Gamma((j+1)\alpha+1)Q_{k-j}\Gamma(\alpha(k-j)+1) = \sum_{j=0}^{k}\binom{k}{j}Q_{j+1}\Gamma((j+1)\alpha+1)X_{k-j}\Gamma(\alpha(k-j)+1),$$

after some algebraic manipulation we obtain

$$\Gamma((k+1)\alpha+1)X_0 Q_{k+1} = n\sum_{j=0}^{k}\frac{k!\Gamma(\alpha(k-j)+1)\Gamma((j+1)\alpha+1)}{j!(k-j)!}X_{j+1}Q_{k-j}$$
$$-\sum_{j=0}^{k-1}\frac{k!\Gamma((j+1)\alpha+1)\Gamma(\alpha(k-j)+1)}{j!(k-j)!}X_{k-j}Q_{j+1}$$

Let $i = j + 1$ in the first sum and $i = k - j$ in the second sum, then we get

$$\Gamma((k+1)\alpha+1)X_0 Q_{k+1} = n\sum_{i=1}^{k+1}\frac{k!\Gamma(\alpha(k+1-i)+1)\Gamma(i\alpha+1)}{(i-1)!(k+1-i)!}X_i Q_{k+1-i}$$
$$-\sum_{i=1}^{k}\frac{k!\Gamma((k+1-i)\alpha+1)\Gamma(i\alpha+1)}{(k-i)!i!}X_i Q_{k+1-i},$$

if $m = k+1$, then the last equation become

$$\Gamma(m\alpha+1)X_0 Q_m = n\sum_{i=1}^{m} \frac{(m-1)!\Gamma(\alpha(m-i)+1)\Gamma(i\alpha+1)}{(i-1)!(m-i)!} X_i Q_{m-i}$$
$$-\sum_{i=1}^{m-1} \frac{(m-1)!\Gamma((m-i)\alpha+1)\Gamma(i\alpha+1)}{(m-1-i)!i!} X_i Q_{m-i}$$

or

$$\Gamma(m\alpha+1)X_0 Q_m = n\sum_{i=1}^{m} \frac{(m-1)!\Gamma(\alpha(m-i)+1)\Gamma(i\alpha+1)}{(i-1)!(m-i)!} X_i Q_{m-i}$$
$$-\sum_{i=1}^{m-1} \frac{(m-1)!(m-i)\alpha\Gamma((m-i)\alpha)\Gamma(i\alpha+1)}{(m-1-i)!i!} X_i Q_{m-i}.$$

By adding the zero value $\left\{-\frac{(m-m)\alpha\,\Gamma(m\alpha+1)}{m} X_m Q_0\right\}$ to the second sum of the last equation,

we get

$$\Gamma(m\alpha+1)X_0 Q_m = n\sum_{i=1}^{m} \frac{(m-1)!\Gamma(\alpha(m-i)+1)\Gamma(i\alpha+1)}{(i-1)!(m-i)!} X_i Q_{m-i}$$
$$-\sum_{i=1}^{m} \frac{(m-1)!\Gamma((m-i)\alpha+1)\Gamma(i\alpha+1)}{(m-1-i)!i!} X_i Q_{m-i},$$

the coefficients $Q_m$ could be determined by

$$Q_m = \frac{1}{\Gamma(m\alpha+1)X_0}\sum_{i=1}^{m} \frac{(m-1)!\Gamma(\alpha(m-i)+1)\Gamma(i\alpha+1)}{i!(m-i)!}[in-m+i]X_i Q_{m-i}, \quad \forall\, m\geq 1.$$

at $m=0$, $Q_0 = X_0^n$, $Q_1 = \frac{\Gamma(\alpha+1)n}{\Gamma(\alpha+1)A_0} X_1 Q_0$.

Putting $n=3$ in the last equation gives

$$Q_m = \frac{1}{\Gamma(m\alpha+1)X_0}\sum_{i=1}^{m} \frac{(m-1)!\Gamma(\alpha(m-i)+1)\Gamma(i\alpha+1)}{i!(m-i)!}[4i-m]X_i Q_{m-i}, \quad \forall\, m\geq 1 \quad (16)$$

and $Q_0 = X_0^3$, $Q_1 = \frac{3X_1 Q_0}{X_0}$, ....

Now by differentiating Equation (12) $\alpha$-times and after some manipulations we have

$$D_t^\alpha x = \sum_{n=1}^{\infty} \frac{\Gamma(n\alpha+1)}{\Gamma(n\alpha+1-\alpha)} X_n T^{n-1},$$

$$D_t^\alpha y = \sum_{n=1}^{\infty} \frac{\Gamma(n\alpha+1)}{\Gamma(n\alpha+1-\alpha)} Y_n T^{n-1},$$

$$D_t^\alpha z = \sum_{n=1}^{\infty} \frac{\Gamma(n\alpha+1)}{\Gamma(n\alpha+1-\alpha)} Z_n T^{n-1}, \quad (17)$$

$$D_t^\alpha r = \sum_{n=1}^{\infty} \frac{\Gamma(n\alpha+1)}{\Gamma(n\alpha+1-\alpha)} R_n T^{n-1},$$

substituting Equations (12) and (17) in Equation (11) for $x$, $y$, $z$ and $r$ respectively, we could determine the coefficients of the power series $X_{n+1}$, $Y_{n+1}$, $Z_{n+1}$ and $R_{n+1}$ from

$$X_{n+1} = -\frac{\Gamma(n\alpha+1)}{\Gamma((n+1)\alpha+1)}\left[3aQ_n + b\sum_{j=0}^{n} X_j Y_{n-j} + c\sum_{j=0}^{n} X_j Z_{n-j}\right],$$

$$Y_{n+1} = \frac{\Gamma(n\alpha+1)}{\Gamma((n+1)\alpha+1)}\left[aQ_n - b\sum_{j=0}^{n} X_j Y_{n-j}\right],$$

$$Z_{n+1} = \frac{\Gamma(n\alpha+1)}{\Gamma((n+1)\alpha+1)}\left[b\sum_{j=0}^{n} X_j Y_{n-j} - c\sum_{j=0}^{n} X_j Z_{n-j}\right],$$

and

$$R_{n+1} = \frac{c\Gamma(n\alpha+1)}{\Gamma((n+1)\alpha+1)}\sum_{j=0}^{n} X_j Z_{n-j}.$$

with the initial values

$$X_0 = 1; Y_0 = 0;\ Z_0 = 0;\ R_0 = 0.$$

### 4. Results

First, we run the MATHEMATICA code to calculate the abundance of each element at $\alpha = 1$. We use six series terms and the time 200. The calculations are performed for the pure helium gas with $X_0 = 1; Y_0 = 0;\ Z_0 = 0;\ R_0 = 0$, the temperature and density are $10^8$ K and $10^4$ gcm$^{-3}$ respectively. Comparison with the numerical solution gives good agreement with maximum relative errors are 0.003, 0.002, 0.00012 and 0.00025 for the x, y,z and r respectively. To check the accuracy of the computations, we investigate the conservation formula $x + y + z + r = 1$. If the sum is not 1, so there is a trouble through the computation, Table 1 lists the absolute errors ($\varepsilon$) at different times. As we see from the table, the absolute error of the sum increases with increasing time and the maximum error $4 \times 10^{-4}$.

To investigate the effects of changing the fraction parameter on the abundance of the four elements, we have performed the calculations at the fractional parameters $\alpha = 0.25, 0.5, 0.75$ and 1 respectively. Figures 5 to 7 illustrate the relation between helium concentration and the product abundances of $C^{12}$, $O^{16}$ and $Ne^{20}$ respectively. When the helium concentration is about 0.7 the difference between the product abundance of $C^{12}$ computed at different values of $\alpha$ is very slight. This differences become larger when the helium abundance goes to zero.

The behavior of the product of the $O^{16}$ is different than that of $C^{12}$, the product abundance computed at $\alpha = 1$ is smaller than that computed at $\alpha$ smaller than one. The situation is reversed as the helium concentration goes from 0.3 to zero, the product of $Ne^{20}$ have the same behavior as the helium concentration less than 0.5, after that the difference become remarkable. Then the fractional parameter could be used to adjust the product abundance of the synthases elements.

Table 1: The absolute errors of the sum of product abundances.

| Time | $\varepsilon$ | | | |
| --- | --- | --- | --- | --- |
| | $\alpha = 1$ | $\alpha = 0.75$ | $\alpha = 0.5$ | $\alpha = 0.25$ |
| 0 | $1.2 \times 10^{-7}$ | $1.0 \times 10^{-7}$ | $2.0 \times 10^{-7}$ | $2.0 \times 10^{-7}$ |
| 500 | $4.5 \times 10^{-5}$ | $5.6 \times 10^{-5}$ | $4.7 \times 10^{-5}$ | $4.9 \times 10^{-5}$ |
| 1000 | $9.4 \times 10^{-5}$ | $1.3 \times 10^{-4}$ | $1.7 \times 10^{-4}$ | $1.9 \times 10^{-4}$ |
| 1500 | $1.4 \times 10^{-4}$ | $1.9 \times 10^{-4}$ | $2.5 \times 10^{-4}$ | $2.9 \times 10^{-4}$ |
| 2000 | $1.9 \times 10^{-4}$ | $2.6 \times 10^{-4}$ | $3.3 \times 10^{-4}$ | $3.8 \times 10^{-4}$ |

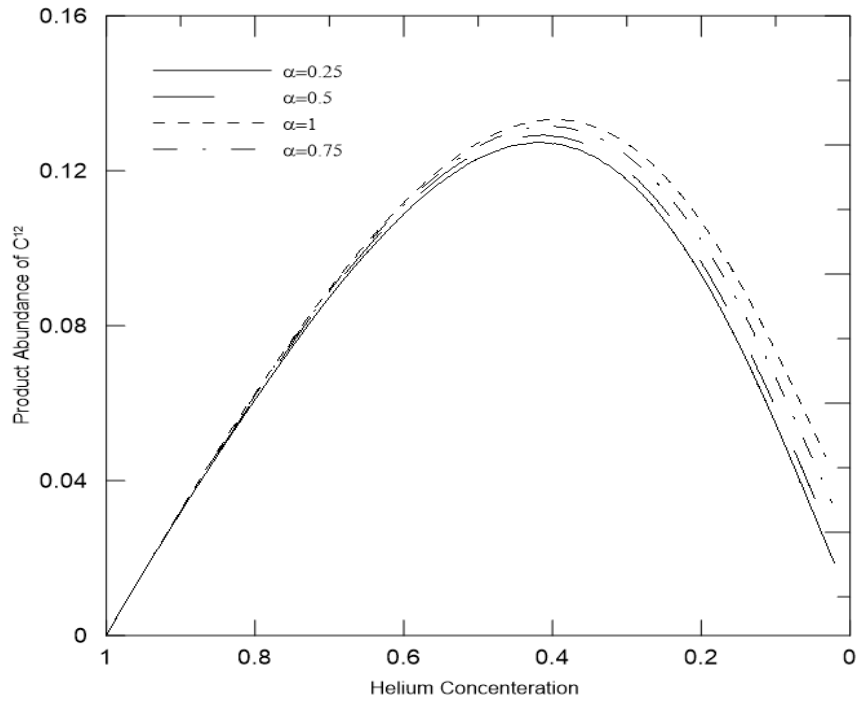

**Figure 1:** The helium concentration as a function of the product abundance of $C^{12}$. The different line shapes indicate different values of the fractional parameter $\alpha$.

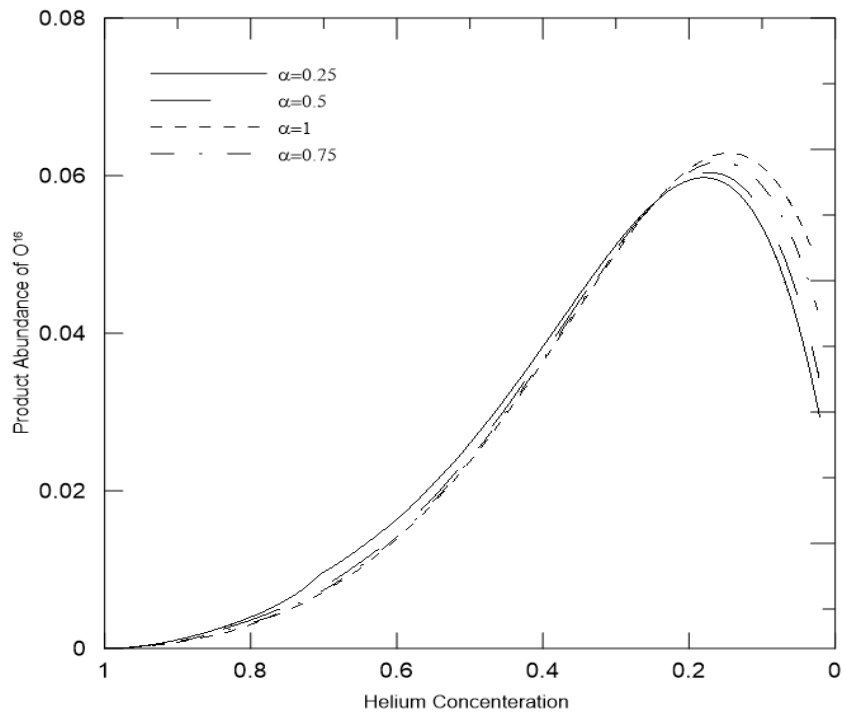

**Figure 2:** The helium concentration as a function of the product abundance of $O^{16}$. The different line shapes indicate different values of the fractional parameter $\alpha$.

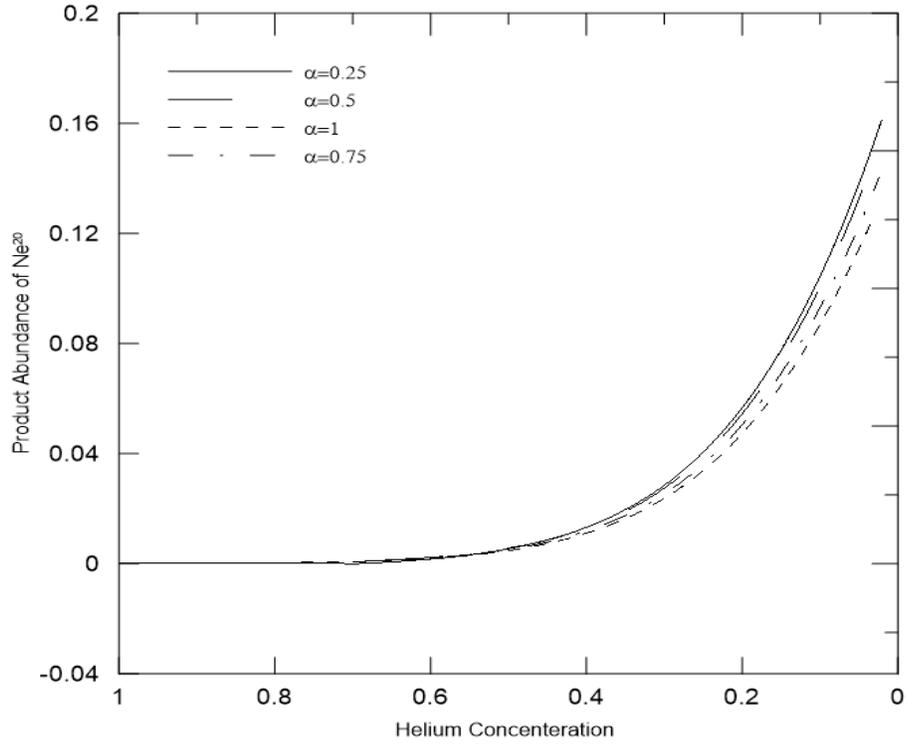

Figure 3: The helium concentration as a function of the product abundance of $^{20}$Ne. The different line shapes indicate different values of the fractional parameter $\alpha$.

## 5. Conclusions

In summarizing the present paper, analytical solution to the abundances differential equations of the helium burning network in its fractional form has been performed. We constructed recurrence relations for the power series coefficients of product abundance. Symbolic as well as numerical computations at $\alpha = 1$ (the integer version of the network) are computed and compared. The maximum relative error is $\varepsilon = 0.003$. The effects of the fractional parameter on the product abundances has been investigated, we found that the product abundance is affected by changing the fractional parameter. The numerical results showed different behaviors when compared with integer model solutions.

**References**

Clayton, D. D., 1983, Principles of Stellar Evolution and Nucleosynthesis, University of Chicago Press, Chicago.

Nouh, M. I., Sharaf, M. A. and Saad, A. S., 2003, Astron. Nachr., 324, 432.

Duorah, H. L. and Kushwaha, R. S., 1963, Helium-Burning Reaction Products and the Rate of Energy Generation, ApJ, 137, 566.

Hix, W. R.; Thielemann, F.-K., 1999, Computational methods for nucleosynthesis and nuclear energy generation, J. Comput. Appl. Math., 109, 321.

Podlubny, I., 1999, Fractional Differential Equations, Acad. Press, London.

Sokolov, I.M., Klafter, J., Blumen, A., 2002, Phys. Today 55, 48.

Kilbas, A.A., Srivastava, H.M., Trujillo, J.J., 2006, Theory and Applications of Fractional Differential Equations. Elsevier, Amsterdam.

Laskin, N., 2000, Phys. Rev. E 62, 3135.
El-Nabulsi, R.A., 2011, Appl. Math. Comput. 218, 2837.

Bayin, S. S. and Krisch, J. P., 2015, Ap&SS, 359, 58.

Abdel-Salam, E. A-B., and Nouh, M. I., 2016, Astrophysics, 59, 398.

Nouh, M. I and Abdel-Salam, E. A., 2017, Iranian Journal of Science and Technology, Transactions A: Science, in press.

**Appendix: Fractional Calculus**

The modified Riemann–Liouville derivative is written as (Jumarie, 2010)

$$D_x^\alpha f(x) = \begin{cases} \dfrac{1}{\Gamma(-\alpha)} \int_0^x (x-\xi)^{-\alpha-1}[f(\xi)-f(0)]d\xi, & \alpha < 0 \\ \dfrac{1}{\Gamma(1-\alpha)} \dfrac{d}{dx} \int_0^x (x-\xi)^{-\alpha}[f(\xi)-f(0)]d\xi, & 0 < \alpha < 1 \\ \dfrac{1}{\Gamma(n-\alpha)} \dfrac{d^n}{dx^n} \int_0^x (x-\xi)^{n-\alpha-1}[f(\xi)-f(0)]d\xi, & n \leq \alpha < n+1, n \geq 1. \end{cases} \quad (1)$$

Five useful formulas of Jumarie's modified Riemann–Liouville derivative are given by

$$D_x^\alpha x^\gamma = \frac{\Gamma(\gamma+1)}{\Gamma(\gamma+1-\alpha)} x^{\gamma-\alpha}, \qquad \gamma > 0, \tag{2}$$

$$D_x^\alpha (c\,f(x)) = c\,D_x^\alpha f(x), \tag{3}$$

$$D_x^\alpha [f(x)g(x)] = g(x)D_x^\alpha f(x) + f(x)D_x^\alpha g(x), \tag{4}$$

$$D_x^\alpha f[g(x)] = f_g'[g(x)]D_x^\alpha g(x), \tag{5}$$

and

$$D_x^\alpha f[g(x)] = D_g^\alpha f[g(x)](g_x')^\alpha, \tag{6}$$

where $c$ is constant. Equations (4) and (6) could be obtained from

$$D_x^\alpha f(x) \cong \Gamma(\alpha+1) D_x f(x). \tag{7}$$

He et al.(2012) has modified the chain rule (Equation (5)) to

$$D_x^\alpha f[g(x)] = \sigma_x f_g'[g(x)]D_x^\alpha g(x), \tag{8}$$

where $\sigma_x$ is called the fractal index determined in terms of gamma functions. Therefore, Equations (4) and (6) will be modified to the following forms

$$D_x^\alpha [f(x)g(x)] = \sigma_x \{g(x)D_x^\alpha f(x) + f(x)D_x^\alpha g(x)\}, \tag{9}$$

and

$$D_x^\alpha f[g(x)] = \sigma_x D_g^\alpha f[g(x)](g_x')^\alpha. \tag{10}$$

Equations (9) and (10) are used to solve the system of differential equations introduced in the present paper.